\documentstyle[12pt]{article}

\def\beq{\begin{equation}}
\def\eeq{\end{equation}}
\def\beqa{\begin{eqnarray}}
\def\eeqa{\end{eqnarray}}

\begin{document}

\begin{flushright}
ITP-SB-97-42
\end{flushright}

\vspace{3mm}
\begin{center}
{\Large \bf Threshold Resummation
for QCD Hard Scattering\footnote{Presented
at the Fifth International Workshop on Deep Inelastic
Scattering and QCD, Chicago, IL, April 14-18, 1997.}} 
\end{center}
\vspace{2mm}
\begin{center}
Nikolaos Kidonakis\\
\vspace{2mm}
{\it Department of Physics and Astronomy\\
University of Edinburgh\\
Edinburgh EH9 3JZ, Scotland} \\
\vspace{4mm}
George Sterman \\
\vspace{2mm}
{\it Institute for Theoretical Physics\\
State University of New York at Stony Brook\\
Stony Brook, NY 11794-3840, USA} \\  
\vspace{2mm}
April 1997
\end{center}

\begin{abstract}
We describe the resummation of singular distributions at partonic
threshold in heavy quark production, generalizing results for
Drell-Yan and other electroweak hard-scattering processes.
\end{abstract}

\section{Introduction}

The normalization of the inclusive cross section
for heavy quark production is of great
interest for the interpretation of top production \cite{LSvN}.
In this and related cross sections, it is desirable to resum certain
higher-order effects to all orders in perturbation 
theory.  To see what effects are involved in such resummations,
and why they are of interest, we recall the general factorized form
of heavy quark production cross sections in hadron-hadron collisions,
which may be written schematically as
\beq
{d\sigma_{h_1h_2\rightarrow Q{\bar Q}}\over dQ^2 dy d\cos\theta}
=
\sum_i
\phi_{i/h_1}\otimes\hat{\sigma}_{i{\bar i}\rightarrow Q{\bar Q}} \otimes \phi_{{\bar i}/h_2}\, ,
\label{sigmaphifact}
\eeq
where the $\phi$'s are parton distributions, in convolution (denoted $\otimes$)
with a hard-scattering function $\hat\sigma$, which is calculable in perturbation
theory.  $Q^2\ge 4m_Q^2$ is the total invariant mass squared of the
heavy pair, and $y$ is its rapidity.
For our purposes, it is only necessary to consider quark-antiquark
and gluon fusion processes in the sum over parton type $i$.

The particular corrections in which we are interested are the
``plus distributions" in the variable
$1-z\equiv 1-Q^2/x_ax_bS$, with $S=(p_{h_1}+p_{h_2})^2$.
For quark annihilation and gluon fusion, plus distributions
occur in $\hat\sigma$ in the general form
\beq
{\hat \sigma}_{i{\bar i}\rightarrow Q{\bar Q}}(1-z)
\sim 
\cdots+ \left({\alpha_s\over\pi}\right)^n\; \left [ {\ln^{2n-1}(1-z)\over 1-z}\right]_+
+\dots\, .
\eeq
Such corrections can be numerically important
when integrated against realistic parton distributions.

Simple manipulations show that the
 plus distributions in $\hat\sigma_{i{\bar i}\rightarrow Q{\bar Q}}$ 
may be computed from the ratio of the
moments of the {\it purely partonic}
cross section $\sigma_{i{\bar i}\rightarrow Q{\bar Q}}$ with respect to $\tau\equiv Q^2/S$
to the moments of the {\it partonic} parton distributions $\phi_{i/i}(x)$
with respect to $x$,
\beq
\hat\sigma_{i{\bar i}\rightarrow Q{\bar Q}}(N)={\sigma_{i{\bar i}\rightarrow Q{\bar Q}}(N)
\over
\left[ \phi_{i/i}(N)\right]^2}\, ,
\label{firstratio}
\eeq
where for simplicity
we denote the moments $\sigma(N)=\int_0^1d\tau \tau^N\sigma(\tau)$ 
and $\phi(N)=\int_0^1dx x^N\phi(x)$ by
their arguments, rather than by using a new notation for the transformed functions.
In these terms, the plus distribution $[\ln^m(1-z)/1-z]_+$ produces
a logarithmic moment dependence $\ln^{m+1}N$.  
In the Drell-Yan cross section,
based on an electroweak (and  therefore color singlet)
hard scattering,  the $\ln N$-dependence 
is known to exponentiate \cite{oldDY}.

\section{Factorization for QCD Hard Scattering}

We have recently shown how to
treat nonleading logarithms for heavy quark
production \cite{NKGS}.  We employ, as in Drell-Yan, an intermediate
factorization, which separates the cross section into functions
associated with three sorts of quanta:  quanta collinear to the
incoming partons, hard quanta, and soft quanta in the central
region.  The schematic form of this factorization is
\beq
\sigma_{i{\bar i}\rightarrow Q{\bar Q}}
=
\psi_{i/i}\; \otimes\; h_J^*h_I\; \otimes\; S_{JI}\; \otimes\; \psi_{{\bar i}/{\bar i}}\, ,
\label{psifact}
\eeq
where the convolution is now in terms of energy in the center of 
mass frame of the annihilating partons.
In Eq.\ (\ref{psifact}), collinear quanta are organized into the $\psi$'s, hard quanta into the
$h$'s and soft into $S_{JI}$.  The indices $J$ and $I$ refer to the
color content of the hard scattering.  For example, in the process
$q{\bar q}\rightarrow Q{\bar Q}$, the hard scattering is a linear
combination of color octet and color singlet exchange between the
light quark $q$ and the heavy quark $Q$. The soft function $S_{JI}$ 
summarizes soft radiation due to a hard scattering with color exchange
$I$ in the amplitude, and $J$ in the complex conjugate amplitude, noting
that interference between different color exchanges is possible.  
Moments of the 
hard scattering function $\hat{\sigma}(N)$ in heavy quark production may be found by
comparing the moments of (\ref{psifact}) with (\ref{firstratio}),
\beq
\hat{\sigma}_{i{\bar i}\rightarrow Q{\bar Q}}(N)
=
\left [ \psi_{i/i}(N)\over \phi_{i/i}(N)\right]^2 S_{JI}(N)h^*_Jh_I\, .
\label{secondratio}
\eeq
The advantage of this form over (\ref{firstratio}) is that
the functions $\psi$ and $\phi$ are universal (and have been
analyzed in the context of Drell-Yan), and that the effects of
soft gluons are factorized.  

\section{Heavy Quarks from Light Quarks}

The separation of virtual soft gluons between hard and soft
functions depends on a new factorization scale.  Such
ambiguities are familiar in the normal factorization of collinear
singularities.  At the same time, the soft function,
$S_{JI}(Q/N\mu,\alpha_s(\mu))$, contains infrared
and ultraviolet logarithms only.  The moment dependence
of $S_{JI}$ is therefore controlled by a renormalization group equation, in
terms of a matrix $\Gamma_{IK}$ of anomalous dimensions.  The rank of $\Gamma_{IK}$
is simply the number of basis color exchanges (singlet, octet, etc.) 
necessary to describe the hard scattering.   In heavy quark
production, $\Gamma_{IK}$ 
depends on the directions and velocities of the outgoing pair.  For example,
consider the process
\beq
q(p_q)+{\bar q}(p_{\bar q})\rightarrow Q(p_Q)+{\bar Q}(p_{\bar Q})\, ,
\eeq
and define $t_1=(p_q-p_Q)^2-m_Q^2$, $u_1=(p_{\bar q} - p_{Q})^2-m_Q^2$.
In a color basis in which the hard scatterings describe $t$-channel 
singlet (I=1) and octet (I=2) exchange, the anomalous dimension matrix is given
by \cite{NKGS}
\begin{eqnarray}
\Gamma_{11}&=&-\frac{\alpha_s}{\pi}C_F[L_{\beta}+1+\pi i],\
\Gamma_{21}=\frac{2\alpha_s}{\pi}
\ln\left(\frac{t_1}{u_1}\right),\ 
\Gamma_{12}=\frac{\alpha_s}{\pi}
\frac{C_F}{C_A} \ln\left(\frac{t_1}{u_1}\right),\nonumber\\
\Gamma_{22}&=&\frac{\alpha_s}{\pi}\left\{C_F
\left[4\ln\left(\frac{t_1}{u_1}\right)
-L_{\beta}-1-\pi i\right]\right.
\nonumber \\ &&
\left.+\frac{C_A}{2}\left[-3\ln\left(\frac{t_1}{u_1}\right)
-\ln\left(\frac{m^2s}{u_1t_1}\right)+L_{\beta}+\pi i \right]\right\}\, .
\label{Gammaoneeight}
\end{eqnarray}
where the function $L_\beta$ includes the exchange of
a gluon between the outgoing, heavy pair,
\begin{equation}
L_{\beta}=\frac{1-2m^2/s}{\beta}\left(\ln\frac{1-\beta}{1+\beta}
+\pi i \right)\, .
\end{equation}
The quantity $\beta=\sqrt{1-4m^2/Q^2}$ is the center-of-mass velocity
of the produced pair.
We note that the particular form shown in Eq.\ (\ref{Gammaoneeight})
depends on the gauge in which the
functions $\psi$ are constructed.  In this case, we
have chosen $A^0=0$ gauge.  Gauge dependence appears only in
the diagonal elements of $\Gamma_{JI}$, and is compensated by the
$\psi$'s of Eq.\ (\ref{psifact}).

Because $\Gamma_{JI}$ is generally nondiagonal, the
logarithms of moments exponentiate simply at next-to-leading
logarithms only in a basis in which it is diagonal. 
In this case, moments of the resummed cross section 
are proportional to an overall factor,  whose $N$
dependence is identical to that of the Drell-Yan cross section \cite{oldDY},
and which includes the leading logarithms, times 
a color-dependent factor,
\begin{eqnarray}
{\hat \sigma}_{f{\bar f}\rightarrow Q{\bar Q}}(N) &=&
 A'(\alpha_s(Q^2))\; h^*_J\left(\alpha_s(Q^2)\right)\; 
{\tilde S}_{JI} \biggl ( 1,\alpha_s([Q/N]^2) \biggr )\; 
h_I\left(\alpha_s(Q^2)\right) 
\nonumber \\
&\ & 
\quad \quad \times
\exp \left[ E_{JI}(N,\theta,Q^2)\right ] \, .
\label{omegaofn}
\end{eqnarray}
Here $A'$ is an overall constant, while the color-dependent exponents are
\begin{eqnarray}
E_{JI}(N,\theta,Q^2)&=&E_{\rm DY}(N,Q) 
 - \int_0^1 dz \frac{z^{N-1}-1}{1-z}
\biggl[ g_3^{(I)}[\alpha_s((1-z)^2 Q^2),\theta] \nonumber \\
&\ & \quad  +g_3^{(J)*}[\alpha_s((1-z)^2 Q^2),\theta]\, \biggr ]\, ,
\label{Eofn}
\end{eqnarray}
where $E_{\rm DY}$ is the Drell-Yan exponent, while $g_3$ is
given by the eigenvalues $\lambda_I$ of $\Gamma$, shifted
to match the Drell-Yan result,
$g_3^{(I)}[\alpha_s,\theta]=-\lambda_I[\alpha_s,\theta]
+{\alpha_s\over \pi}C_F$.
In this notation, $\theta$ is the center-of-mass scattering
angle, which, along with $\beta$ determines the ratios
of kinematic variables in the matrix (\ref{Gammaoneeight}).

Of course, predictions for physical cross sections require
the inversion of the exponentiated moments, \cite{AMS,recent}.
Overall, however, the most important result is
that nonleading logarithms are under control.
If numerical results \cite{KSV} are large, they constitute substantive
predictions that we may in principle compare to experiment.
If small, they give us added confidence in NLO results.
Finally, let us note that we have computed the corresponding
anomalous dimensions for gluon fusion \cite{NKGS} and scattering \cite{NOS}. 

{\em Acknowledgements}.
This work was supported in part by the National Science Foundation,
under grant PHY9309888 and by the PPARC under grant GR/K54601.
We wish to thank 
Lyndon Alvero, Harry Contopanagos, Gianluca Oderda, Gregory Korchemsky, Eric Laenen,
and Jack Smith for helpful conversations.

\end{document}